\documentclass[final,5p,times,twocolumn]{elsarticle}

\usepackage{amsmath,amssymb}
\usepackage{bm}
\usepackage{booktabs}
\usepackage{graphicx}
\usepackage{multirow}
\usepackage{xcolor}
\usepackage{hyperref}
\usepackage{lineno}

\hypersetup{
  colorlinks=true,
  linkcolor=blue,
  citecolor=blue,
  urlcolor=blue
}

\journal{Physics of the Dark Universe}
\biboptions{sort&compress}

\newcommand{\Neff}{N_{\rm eff}}

\begin{document}

\begin{frontmatter}

\title{Cosmological constraints on extra radiation and early dark energy with lepton asymmetry and varying quark masses}

\author[kut]{Yo Toda\corref{cor1}}
\ead{toda.yo@kochi-tech.ac.jp} 
\affiliation[kut]{organization={School of Data \& Innovation, Kochi University of Technology},
                  city={Kochi},
                  country={Japan}}

\begin{abstract}
We investigate cosmological constraints on extra radiation and axionlike early dark energy (EDE), allowing primordial lepton asymmetry or light-quark-mass variation during big bang nucleosynthesis (BBN). We combine Planck CMB, DESI DR2 BAO, Pantheon+ supernovae, and primordial helium and deuterium measurements. When both extra radiation and lepton asymmetry are allowed, we obtain $N_{\rm eff}=3.27\pm0.16$, $\xi=0.017\pm0.011$, and $H_0=69.24\pm0.92\,{\rm km\,s^{-1}\,Mpc^{-1}}$. A positive lepton asymmetry compensates the helium increase caused by extra radiation, allowing a higher inferred Hubble constant even with the LBT helium measurement, which agrees with standard BBN. For EDE, primordial-abundance measurements instead restrict the higher baryon densities associated with larger EDE fractions, which reduce the predicted deuterium abundance. Allowing lepton asymmetry does not relax this restriction. In contrast, a positive light-quark-mass variation increases deuterium and compensates this reduction, permitting larger EDE fractions and a higher inferred Hubble constant. The same variation leaves the extra-radiation constraint nearly unchanged. We also find that allowing lepton asymmetry relaxes the 95\% upper limit on the summed neutrino mass from $0.0640\,{\rm eV}$ to $0.0777\,{\rm eV}$ in the extra-radiation model.
\end{abstract}


\end{frontmatter}


\section{Introduction}
\label{sec:introduction}
The standard Big Bang cosmology and the $\Lambda$CDM
model have been remarkably successful in explaining
a wide range of cosmological observations.
In particular, the anisotropies of the cosmic microwave
background (CMB) and the agreement between observed
primordial light-element abundances and the predictions
of big bang nucleosynthesis (BBN) provide strong
evidence for our understanding of the early Universe.
Nevertheless, increasingly precise observations have
revealed discrepancies that challenge the standard
cosmological model.
Among them, the Hubble tension, namely the discrepancy
between the Hubble constant inferred from CMB observations
and independently from baryon acoustic oscillation (BAO)
measurements combined with BBN within
the $\Lambda$CDM framework
~\cite{Planck:2018vyg,Rosenberg:2022sdy,
Tristram:2023haj,ACT:2020gnv,
AtacamaCosmologyTelescope:2025blo,
SPT-3G:2025bzu,Schoneberg:2019wmt,Schoneberg:2022ggi}
and that measured using the local distance ladder
~\cite{Riess:2021jrx,Riess:2025chq,
Breuval:2024lsv,Freedman:2024eph,
Li:2025ife,Blakeslee:2021rqi},
remains a major unresolved problem.

Several extensions of the standard cosmological model
have been proposed to alleviate the Hubble tension.
One class of solutions introduces an additional energy
component in the early Universe, which increases the
expansion rate before recombination and reduces the
sound horizon, thereby allowing a higher inferred
Hubble constant.
Examples of this approach include models with additional relativistic energy density
~\cite{Wyman:2013lza,DEramo:2018vss}
and early dark energy (EDE)
~\cite{Poulin:2018dzj,Poulin:2018cxd,
Braglia:2020bym,Agrawal:2019lmo,
Ye:2020btb,Smith:2019ihp,Lin:2019qug,
Niedermann:2019olb,Niedermann:2020dwg,
Seto:2021xua,Freese:2021rjq,
Niedermann:2021vgd,Bella:2026zuk}.
An alternative approach to reducing the sound horizon
is to modify recombination physics rather than introduce
an additional early energy component.
For example, a varying electron mass changes atomic binding
energies and Thomson scattering, shifting the recombination
and baryon-drag epochs and the associated sound horizons
~\cite{Hoshiya:2022ady, Seto:2022xgx, Seto:2024cgo, Toda:2024ncp, Toda:2024uff, Toda:2025dzd, Sekiguchi:2020teg, Sekiguchi:2020igz, Hart:2017ndk, Hart:2019dxi, Chluba:2023xqj, Toda:2025kcq}.

Primordial light-element abundances provide important
constraints on these extensions.
An increase in the effective number of relativistic species,
$N_{\rm eff}$, generally raises the primordial helium mass
fraction, $Y_P$, thereby imposing BBN constraints on
the ability of additional radiation to alleviate the
Hubble tension.
On the other hand, a positive electron-neutrino degeneracy
parameter, $\xi_e$, reduces $Y_P$.
Consequently, allowing $N_{\rm eff}$ and $\xi_e$ to vary
independently permits a compensation between the effects
of additional radiation and lepton asymmetry,
potentially alleviating the Hubble tension while
maintaining consistency with primordial light-element
abundances~\cite{Seto:2021tad}.

A sizable primordial lepton asymmetry is also theoretically
motivated.
Although conventional leptogenesis generally produces
lepton and baryon asymmetries of comparable magnitudes,
several mechanisms have been proposed to generate a
much larger lepton asymmetry while preserving the
observed small baryon asymmetry.
These include Affleck--Dine leptogenesis, scenarios
involving long-lived lepton-number-carrying Q-balls,
and low-scale leptogenesis
\cite{Casas:1997gx,McDonald:1999in,Kawasaki:2002hq,
Kawasaki:2022hvx,Borah:2022uos,Akita:2025zvq}.
Furthermore, models connecting leptogenesis with the
production of additional relativistic particles, such
as majorons, have also been investigated
\cite{Escudero:2021rfi}.
These scenarios provide particle-physics motivation
for considering primordial lepton asymmetry together
with additional radiation.

The effectiveness of this compensation mechanism depends
 on the adopted primordial helium abundance.
The EMPRESS survey reported a relatively low helium
mass fraction,
$Y_P=0.2387^{+0.0036}_{-0.0031}$
\cite{Yanagisawa:2025mgx}.
These relatively low abundances motivated cosmological
analyses with positive lepton asymmetry
\cite{Seto:2023yal, Escudero:2022okz}.
In particular, a previous analysis combining Planck,
DESI DR1 BAO, Pantheon+, and EMPRESS measurements
found that additional radiation and lepton asymmetry
could substantially alleviate the Hubble tension
\cite{Li:2025rjr}.
More recently, however, the LBT $Y_P$ Project reported
$Y_P=0.2458\pm0.0013$, achieving a precision of
approximately $0.5\%$ \cite{Aver:2026dxv} and finding good agreement
with standard BBN predictions
\cite{Yeh:2026pil,Goldstein:2026iuu}.
A recent cosmological analysis incorporating DESI DR2
and comparing the EMPRESS and LBT measurements
further demonstrated the importance of the adopted
helium abundance for constraints on neutrino chemical
potentials~\cite{Ghedini:2026mvj}.
However, that analysis did not independently vary an
additional radiation component.
These developments motivate a reassessment of whether
additional radiation and lepton asymmetry can still
alleviate the Hubble tension while remaining consistent
with the updated precision BBN measurements.

EDE provides another possible mechanism for alleviating
the Hubble tension by increasing the expansion rate before
recombination and reducing the sound horizon.
Although the EDE contribution in the models considered
here is negligible during BBN, primordial light-element
abundances can still impose significant constraints.
A previous study showed that EDE models can favor a
higher baryon density, thereby reducing the predicted
primordial deuterium abundance and worsening agreement
with BBN observations~\cite{Seto:2021xua}.
That analysis adopted the primordial deuterium abundance
$10^5({\rm D/H}){\rm p}=2.527\pm0.030$,
obtained from a 2018 high-precision determination based
on seven metal-poor quasar absorption systems
~\cite{Cooke:2017cwo}.
More recently, the 2026 Particle Data Group (PDG)
compilation of twelve high-precision measurements,
including a 2024 analysis of a metal-poor absorption
system at $z=3.42$ toward the quasar J1332+0052,
reports a slightly lower value,
$10^5({\rm D/H}){\rm p}=2.508\pm0.029$
~\cite{PDG:2026}, which may somewhat weaken the
deuterium-based constraints on EDE.
Furthermore, primordial abundances can be modified
by lepton asymmetry or variations in the light-quark
masses~\cite{Dent:2007zu, Berengut:2013nh}.
These developments motivate a reassessment of the
BBN constraints on EDE using updated observations,
including the possible effects of primordial lepton
asymmetry and variations in the light-quark masses.

In this work, we revisit the cosmological constraints
on early-Universe solutions to the Hubble tension
using updated observations, including precision BBN
measurements.
We also examine whether primordial lepton asymmetry and quark-mass variation
can alleviate the BBN constraints on these models.
Recent DESI DR2 analyses have placed stringent constraints on the summed neutrino mass, $\sum m_\nu$, and examined their dependence on the effective number of relativistic species, $N_{\rm eff}$. Allowing $N_{\rm eff}$ and $\sum m_\nu$ to vary simultaneously was found to modestly relax the neutrino mass bound~\cite{Elbers:2025vlz}. We therefore also investigate how primordial lepton asymmetry affects these constraints when both parameters are allowed to vary.
This paper is organized as follows.
In Sec.~\ref{sec:models}, we describe the cosmological models,
followed by the data and methodology in Sec.~\ref{sec:data}.
We present our results in Sec.~\ref{sec:results} and discuss
their implications in Sec.~\ref{sec:discussion}.
Finally, Sec.~\ref{sec:conclusions} is devoted to our conclusions.

\section{Cosmological models}
\label{sec:models}

We consider extensions of flat $\Lambda$CDM involving extra relativistic energy density, a primordial neutrino asymmetry, the summed active-neutrino mass, and axionlike early dark energy (EDE). Unless stated otherwise, we fix $\sum m_\nu=0.06\,{\rm eV}$, represented by one massive and two massless active-neutrino eigenstates in CAMB.

\subsection{Extra radiation and lepton asymmetry}

For each active flavor $\alpha=e,\mu,\tau$, the degeneracy
parameter is $\xi_\alpha\equiv\mu_{\nu_\alpha}/T_\nu$.
We assume a common value, $\xi_e=\xi_\mu=\xi_\tau\equiv\xi$,
with distributions
\begin{equation}
f_{\nu_\alpha}(p)=\frac{1}{e^{p/T_\nu-\xi}+1},
\qquad
f_{\bar{\nu}_\alpha}(p)=\frac{1}{e^{p/T_\nu+\xi}+1}.
\end{equation}
The opposite signs of the chemical potentials for neutrinos
and antineutrinos generate a difference in their number
densities. The lepton asymmetry per flavor is given by
\begin{equation}
L_\alpha\equiv\frac{n_{\nu_\alpha}-n_{\bar{\nu}_\alpha}}{n_\gamma}
=\frac{(T_\nu/T_\gamma)^3}{12\zeta(3)}
\left(\pi^2\xi+\xi^3\right).
\end{equation}
For small $|\xi|$, the number-density asymmetry is
linear in $\xi$.
In contrast, the combined energy density of neutrinos and
antineutrinos is invariant under $\xi\rightarrow-\xi$,
so its leading correction is quadratic in $\xi$.
We write the total relativistic density as
\begin{equation}
\rho_{\rm rad}
=\rho_\gamma\left[
1+\frac{7}{8}\left(\frac{4}{11}\right)^{4/3}N_{\rm eff}
\right],
\end{equation}
with $N_{\rm eff}^{\rm SM}=3.044$~\cite{Bennett:2020zkv}.
A common $\xi$ adds
\begin{equation}
\Delta N_{\rm eff}^{\xi}
\simeq
3\left[
\frac{30}{7}\left(\frac{\xi}{\pi}\right)^2
+\frac{15}{7}\left(\frac{\xi}{\pi}\right)^4
\right]
+0.0102\xi^2,
\end{equation}
where the final term accounts for the $\xi$-dependent correction
from non-instantaneous neutrino decoupling~\cite{Li:2024gzf};
the $\xi$-independent correction is already included in
$N_{\rm eff}^{\rm SM}=3.044$.
In the $N_{\rm eff}+\xi$ model, we sample the total
$N_{\rm eff}$ independently of $\xi$, with the contribution
$\Delta N_{\rm eff}^{\xi}$ included in the total radiation density.

These effects of $\xi$ play different roles during BBN.
The change in the radiation energy density modifies the
expansion rate, whereas the asymmetry between electron
neutrinos and antineutrinos directly affects the
charged-current weak reactions that interconvert
neutrons and protons.
The latter effect enters at first order in $\xi_e$.
In equilibrium, the neutron-to-proton ratio is approximately
\begin{equation}
\left(\frac{n}{p}\right)_{\rm eq}
\simeq
\exp\left(-\frac{\Delta m_{np}}{T}-\xi_e\right),
\end{equation}
where $\Delta m_{np}\simeq 1.293$ MeV is the neutron--proton mass difference.
Positive $\xi_e$ therefore lowers the predicted primordial
helium abundance and can partly compensate the helium
increase induced by extra radiation.
Deuterium provides an additional constraint on this
compensation~\cite{Seto:2021xua,Escudero:2022okz}.

We extend $N_{\rm eff}+\xi$ by sampling $\sum m_\nu$ independently, retaining the same radiation and BBN prescriptions. Following the other chains, the total mass is assigned to one massive active-neutrino species. Massive neutrinos suppress late-time structure growth through free streaming and change the CMB-lensing and distance predictions~\cite{Planck:2018vyg}. 

\subsection{Axionlike early dark energy}
\label{sec:ede}

We also study the axionlike EDE scenario~\cite{Poulin:2018cxd,Poulin:2018dzj}, characterized by the potential
\begin{equation}
V(\phi)=m^2f^2\left[1-\cos(\phi/f)\right]^n.
\label{eq:axionpotential}
\end{equation}
The field is initially frozen and subsequently enters an oscillatory regime.
We describe this evolution using the CAMB \texttt{AxionEffectiveFluid}
implementation, fixing $n=3$ and the initial angle
$\theta_i\equiv\phi_i/f=2.8$, following Ref.~\cite{Bella:2026zuk}.
The corresponding averaged equation of state in the oscillatory
regime is $w_n=(n-1)/(n+1)=1/2$.
We denote the transition redshift by $z_t$ and define
the fractional EDE density at this redshift as
\begin{equation}
f_{\rm EDE}\equiv\frac{\rho_{\rm EDE}(z_t)}{\rho_{\rm tot}(z_t)}.
\label{eq:fede_transition}
\end{equation}
We adopt uniform linear priors $10^{-6}\leq f_{\rm EDE}\leq0.20$
and $1000\leq z_t\leq50000$.

Lepton asymmetry is also incorporated into the EDE model
through a common neutrino degeneracy parameter $\xi$.
In the resulting EDE+$\xi$ model, $N_{\rm eff}$ is
determined by $N_{\rm eff}=3.044+\Delta N_{\rm eff}^{\xi}$
rather than independently varied.
While $\xi$ directly modifies neutron--proton conversion
during BBN, the EDE component itself is negligible at
MeV temperatures.
The effect of EDE on primordial light-element abundances
is therefore primarily indirect, through the baryon density
favored by the cosmological data~\cite{Seto:2021xua}.

\subsection{Light-quark-mass variation during BBN}
\label{sec:mq}
We also examine the possibility that the average light-quark mass,
$m_q=(m_u+m_d)/2$, at the time of BBN differs from its present value.
We introduce the parameter
\begin{equation}
\delta_q\equiv\ln\left(\frac{m_q^{\rm BBN}}{m_q^0}\right)
\simeq\frac{m_q^{\rm BBN}-m_q^0}{m_q^0},
	\label{eq:deltaq}
\end{equation}
where the last expression holds for small variations.
A change in $m_q$ modifies nuclear binding energies and nucleon--nucleon
scattering lengths through its effect on the nuclear interaction, thereby
altering the primordial abundances of light elements~\cite{Berengut:2013nh}.
We treat $\delta_q$ as an independent phenomenological parameter
in both the $N_{\rm eff}+\delta_q$ and EDE+$\delta_q$ models.
Its effects are incorporated through the primordial light-element
abundances, while the remaining cosmological evolution follows
the corresponding extra-radiation or EDE prescription described above.


\section{Data and methodology}
\label{sec:data}

We perform Markov chain Monte Carlo analyses using Cobaya~\cite{Torrado:2020dgo},
with theoretical CMB spectra computed using CAMB~\cite{Lewis:1999bs,Lewis:2026mif}.
Our six standard parameters are
$\{\Omega_bh^2,\Omega_ch^2,100\theta_{\rm MC},\tau,\ln(10^{10}A_s),n_s\}$.
Unless otherwise specified, we fix $\sum m_\nu=0.06\,{\rm eV}$,
represented by one massive active-neutrino species.
For the additional parameters, we adopt uniform priors
\begin{align*}
 2&\leq N_{\rm eff}\leq4, &
 -0.1&\leq\xi\leq0.1,\\
 0&\leq\sum m_\nu\leq1\,{\rm eV}, &
 -0.05&\leq\delta_q\leq0.05.
\end{align*}
The EDE parameters and their priors are specified in
Sec.~\ref{sec:ede}.
Our MCMC convergence criterion is $R-1<0.02$.

We use measurements of the CMB, BAO, SNe Ia, and primordial
light-element abundances.
For the CMB, we combine the Planck 2018 low-multipole temperature
and $E$-mode polarization likelihoods~\cite{Planck:2019nip},
the high-multipole NPIPE CamSpec TTTEEE likelihood~\cite{Rosenberg:2022sdy},
and the Planck PR4 lensing likelihood~\cite{Carron:2022eyg}.
For BAO, we use the DESI DR2 measurements~\cite{DESI:2025zgx}.
For SNe Ia, we use the Pantheon+ light-curve sample~\cite{Brout:2022vxf},
which constrains relative luminosity distances.

We impose independent Gaussian likelihoods for the primordial
helium and deuterium abundances.
For helium, we adopt $Y_P=0.2458\pm0.0013$ from the
LBT $Y_p$ Project~\cite{Aver:2026dxv}.
For deuterium, we use $10^5(\mathrm{D/H})=2.508\pm0.029$,
as recommended by the 2026 Particle Data Group review~\cite{PDG:2026}.
This value is based on 12 high-redshift, low-metallicity quasar
absorption systems~\cite{Cooke:2013cba,Cooke:2016rky,Cooke:2017cwo,Riemer-Sorensen:2014aoa,Riemer-Sorensen:2017pey,Balashev:2015hoe,Zavarygin:2017cov,Guarneri:2024qxi,Kislitsyn:2024jvk}.
Unless otherwise stated, all analyses use the same CMB,
DESI DR2, Pantheon+, and primordial-abundance likelihoods.

The baseline BBN abundances are obtained from the PRIMAT
calculations~\cite{Pitrou:2018cgg}, using the 2024 interpolation
table (\texttt{PRIMAT\_Yp\_DH\_ErrorMC\_2024.dat}) provided with CAMB.
The predictions are evaluated at each sampled
$(\omega_b,N_{\rm eff})$, with $\omega_b\equiv\Omega_bh^2$.
We adopt the theoretical uncertainties from the same table,
fixed at the reference cosmology,
$\sigma_{Y_P}^{\rm th}=0.00010911$ and
$\sigma_{10^5\mathrm{D/H}}^{\rm th}=0.027541$,
and add them in quadrature to the observational uncertainties.
The resulting contributions are
\begin{align}
 \chi^2_{Y_P}
 &=\frac{\left(Y_P^{\rm BBN}-0.2458\right)^2}
 {0.0013^2+0.00010911^2},\\
 \chi^2_{\rm D/H}
 &=\frac{\left[10^5(\mathrm{D/H})^{\rm BBN}-2.508\right]^2}
 {0.029^2+0.027541^2}.
\end{align}

Following the treatment of primordial lepton asymmetry in
Ref.~\cite{Escudero:2022okz}, we include the BBN effect of a
nonzero electron-neutrino chemical potential through its
modification of the charged-current weak reactions that
interconvert neutrons and protons.
This changes the neutron-to-proton ratio before nucleosynthesis
and thereby shifts the primordial helium abundance, while also
inducing a smaller change in the deuterium abundance.
The accompanying change in the relativistic neutrino energy
density is included through the total $N_{\rm eff}$ used in
the cosmological model.
For the direct weak-rate effect, the numerical BBN results
can be represented by the fitting relations~\cite{Escudero:2022okz,Pitrou:2018cgg}
\begin{align}
 Y_P^{\rm BBN}(\omega_b,N_{\rm eff},\xi)
 &=Y_P^{(0)}(\omega_b,N_{\rm eff})\,e^{-0.96\xi},\\
 (\mathrm{D/H})^{\rm BBN}(\omega_b,N_{\rm eff},\xi)
 &=(\mathrm{D/H})^{(0)}(\omega_b,N_{\rm eff})\,e^{-0.53\xi},
\end{align}
where $Y_P^{(0)}$ and $(\mathrm{D/H})^{(0)}$ denote the
PRIMAT-based predictions at $\xi=0$ for the same
$(\omega_b,N_{\rm eff})$.
We modify the BBN-consistency routines in CAMB to apply
these correction factors to the baseline BBN predictions.
The corrected helium abundance is also used in the CMB
recombination calculation.

We incorporate the dependence of the primordial abundances on
$\delta_q$ using the BBN response to changes in nuclear binding
energies and scattering lengths calculated in
Ref.~\cite{Berengut:2013nh}.
The logarithmic sensitivities reported in Table VIII of that study are
$\partial\ln Y_P/\partial\ln m_q=-0.56\pm0.34$ and
$\partial\ln(\mathrm{D/H})/\partial\ln m_q=3.9\pm3.4$.
Using their central estimates, we approximate the BBN predictions by
\begin{align}
 Y_P^{\rm BBN}(\omega_b,N_{\rm eff},\delta_q)
 &=Y_P^{(0)}(\omega_b,N_{\rm eff})\,e^{-0.56\delta_q},
 \label{eq:mqYp}\\
 (\mathrm{D/H})^{\rm BBN}(\omega_b,N_{\rm eff},\delta_q)
 &=(\mathrm{D/H})^{(0)}(\omega_b,N_{\rm eff})\,e^{3.9\delta_q}.
 \label{eq:mqDH}
\end{align}
These corrections are implemented in CAMB in the same manner
as those for $\xi$.
The relations approximate the nuclear-response effects for small
$\delta_q$ and do not include possible model-dependent changes
in the neutron lifetime.
We adopt the central response coefficients without marginalizing
over their theoretical uncertainties.
The resulting constraints on $\delta_q$ are therefore conditional
on the adopted nuclear response, particularly the uncertain
deuterium sensitivity, which is not accounted for by the
standard-BBN theoretical errors.


\section{Results}
\label{sec:results}
\begin{figure*}[p]
\centering
\includegraphics[width=0.96\textwidth]{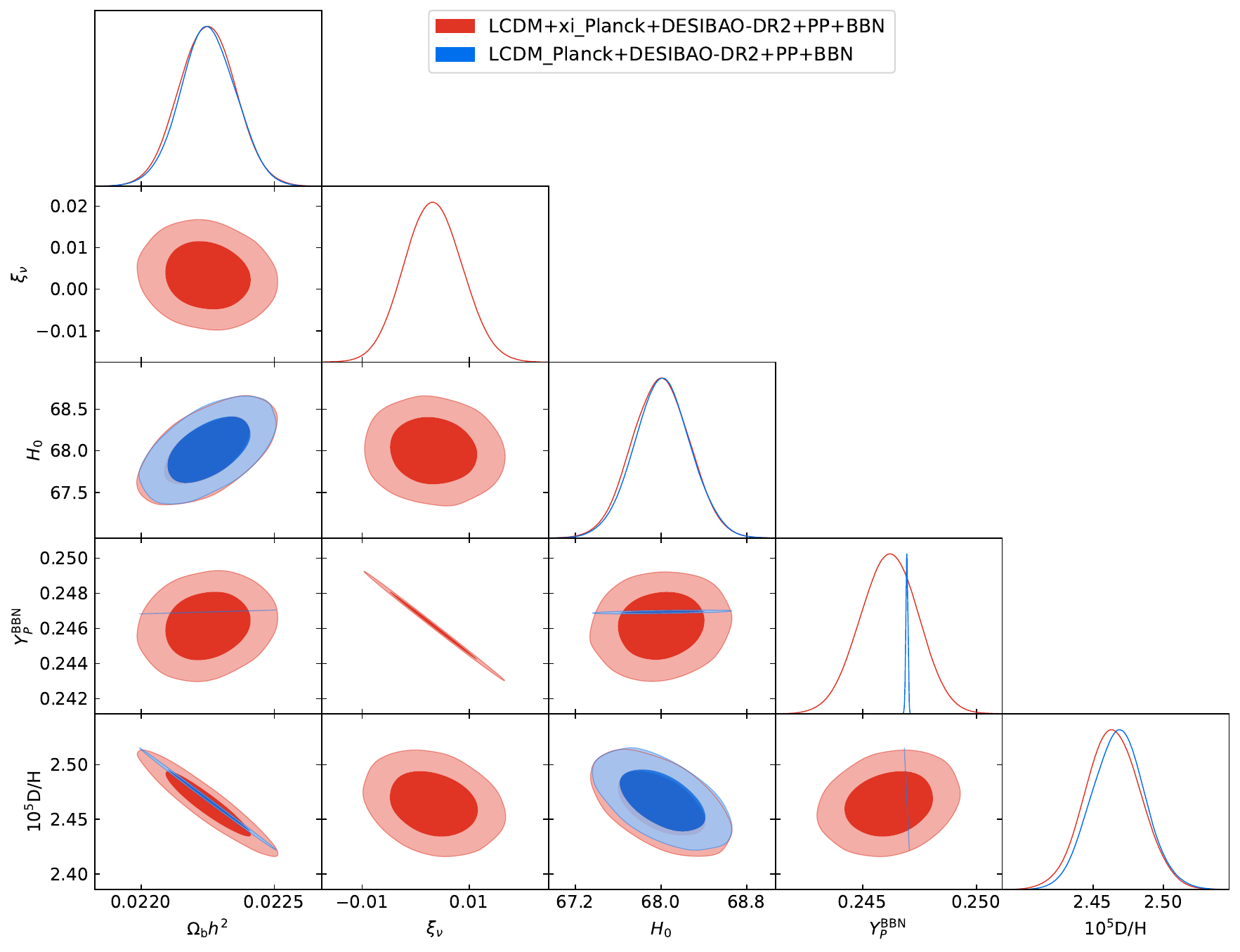}
\caption{Marginalized one- and two-dimensional constraints for
$\Lambda$CDM (blue) and $\Lambda$CDM+$\xi$ (red), using the Planck,
DESI DR2, Pantheon+, and BBN likelihoods.
Allowing lepton asymmetry broadens the predicted primordial
helium abundance while leaving the inferred $H_0$ nearly unchanged.}
\label{fig:lcdmxi}
\end{figure*}

\begin{figure*}[p]
\centering
\includegraphics[width=0.96\textwidth]{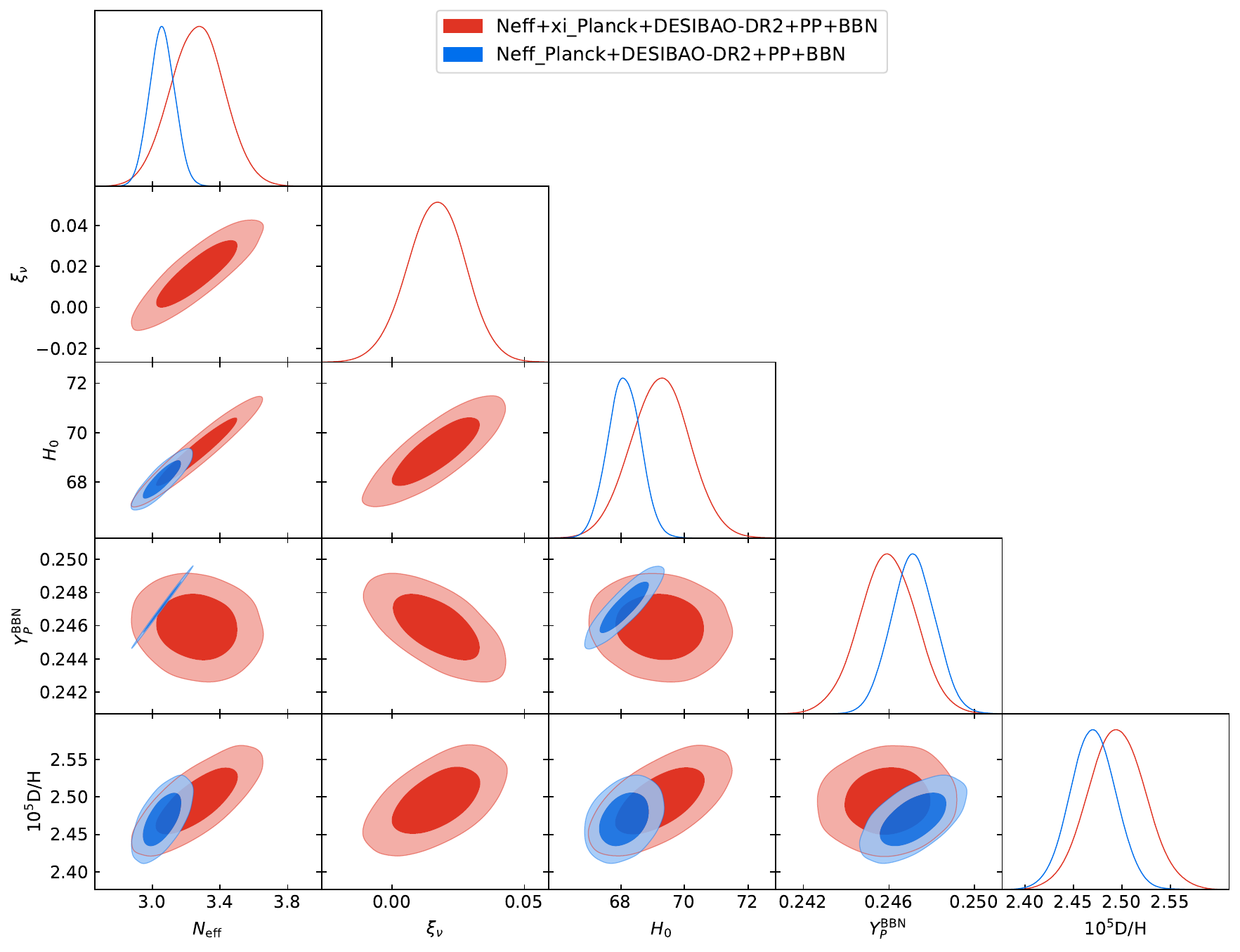}
\caption{Marginalized one- and two-dimensional constraints for
$N_{\rm eff}$ (blue) and $N_{\rm eff}+\xi$ (red), using the Planck,
DESI DR2, Pantheon+, and BBN likelihoods.
Allowing positive lepton asymmetry permits larger values of
$N_{\rm eff}$ and $H_0$ while maintaining consistency with
the primordial helium abundance.}
\label{fig:neffxi}
\end{figure*}

\begin{figure*}[p]
\centering
\includegraphics[width=0.96\textwidth]{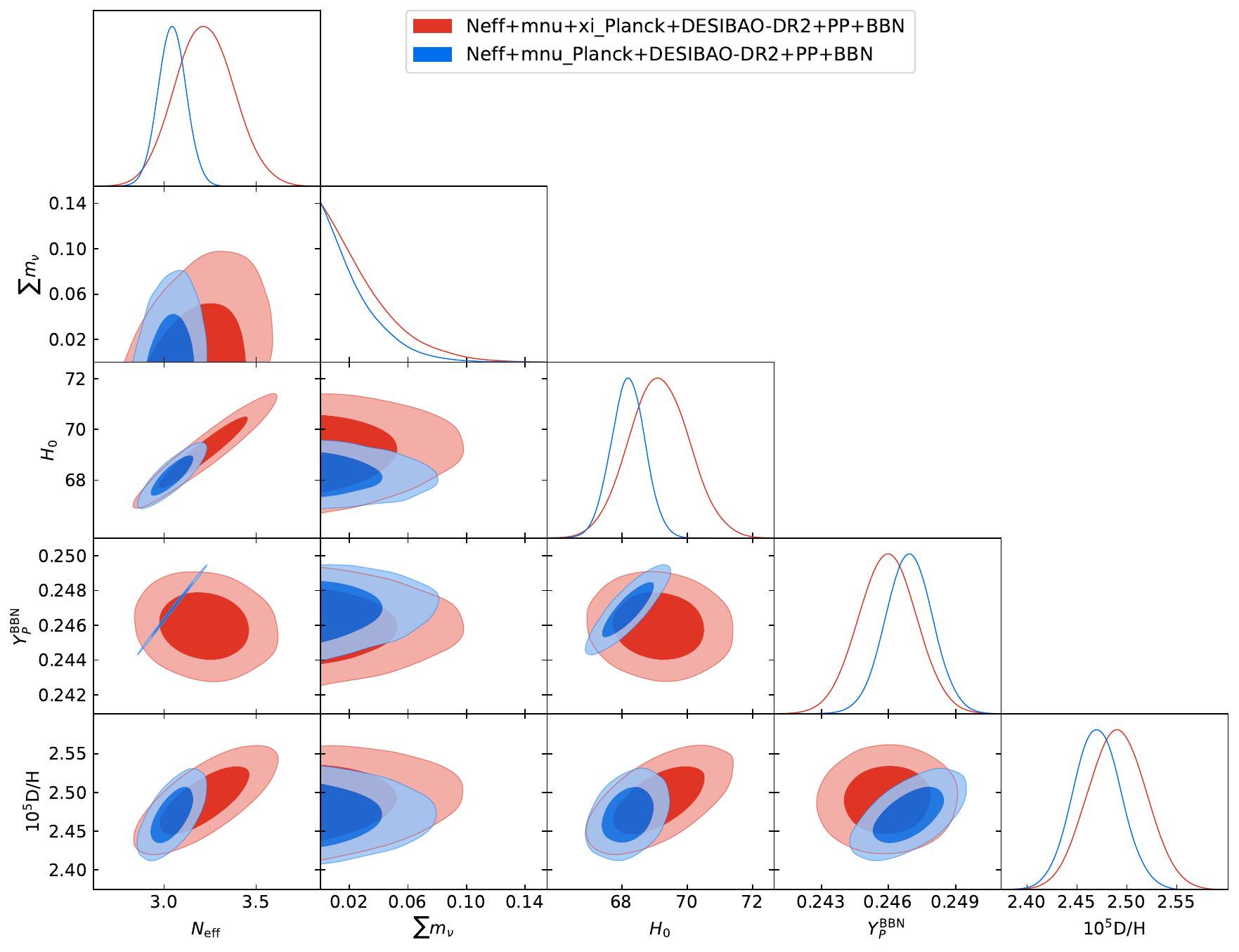}
\caption{Marginalized one- and two-dimensional constraints for
$N_{\rm eff}+\sum m_\nu$ (blue) and
$N_{\rm eff}+\xi+\sum m_\nu$ (red), using the Planck,
DESI DR2, Pantheon+, and BBN likelihoods. The mass posteriors
favor small values, while allowing $\xi$ to vary shifts the
preferred $N_{\rm eff}$ and $H_0$ upward.}
\label{fig:neffmnu}
\end{figure*}

\begin{figure*}[p]
\centering
\includegraphics[width=0.96\textwidth]{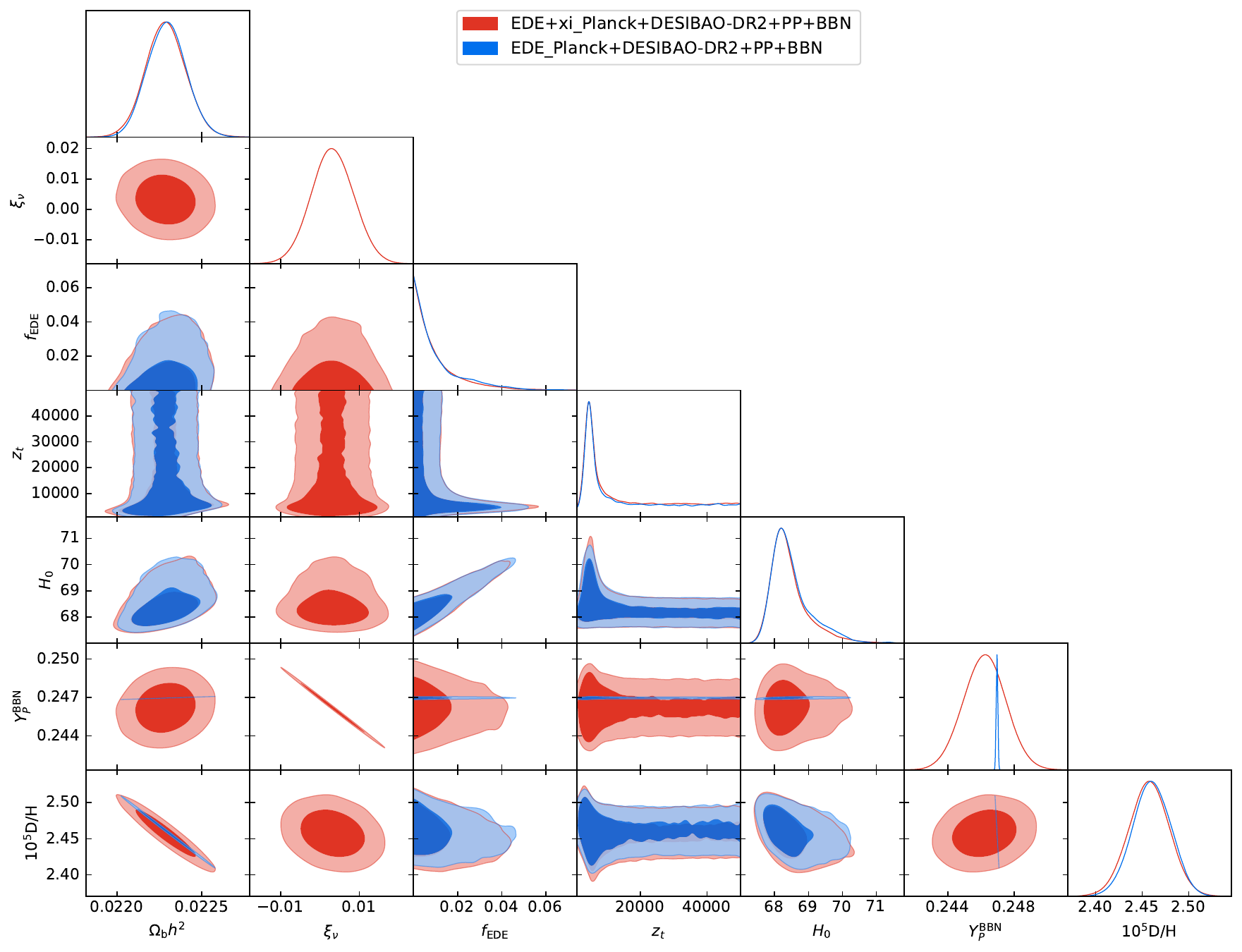}
\caption{Marginalized one- and two-dimensional constraints for
EDE (blue) and EDE+$\xi$ (red), using the Planck,
DESI DR2, Pantheon+, and BBN likelihoods.
Both models favor small EDE fractions, while allowing
lepton asymmetry has little effect on $f_{\rm EDE}$ or
the inferred $H_0$.}
\label{fig:ede}
\end{figure*}

\begin{figure*}[p]
\centering
\includegraphics[width=0.96\textwidth]{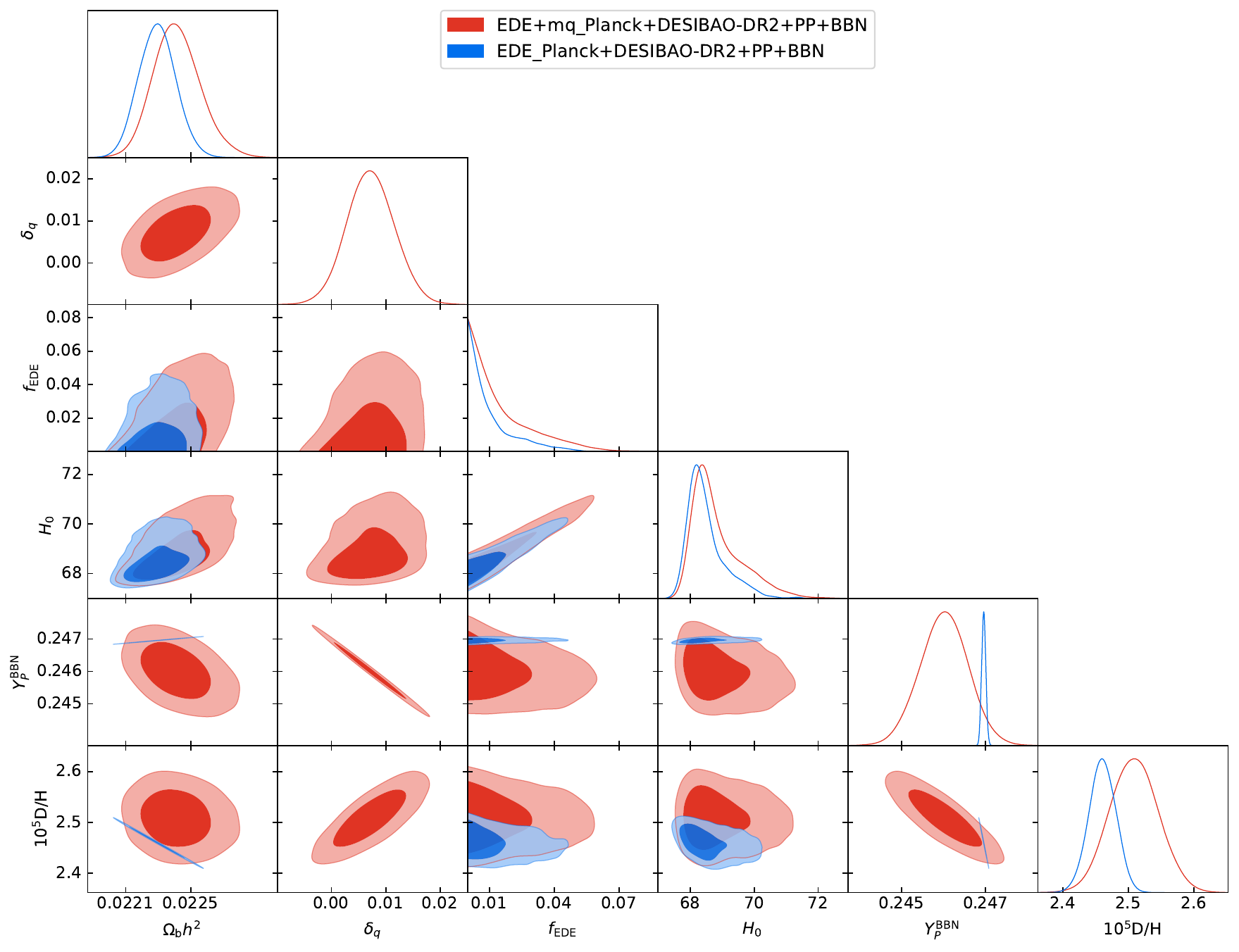}
\caption{Marginalized one- and two-dimensional constraints for
EDE (blue) and EDE+$\delta_q$ (red), using Planck,
DESI DR2 BAO, Pantheon+, and BBN.
Allowing light-quark-mass variation broadens the allowed ranges
of $f_{\rm EDE}$ and $H_0$ and increases the preferred deuterium
abundance despite a higher baryon density.}
\label{fig:edemq}
\end{figure*}

The marginalized constraints for all ten models are summarized in Table~\ref{tab:constraints}. Figs.~\ref{fig:lcdmxi}--\ref{fig:edemq} show posteriors for the principal model comparisons. Unless otherwise stated, uncertainties refer to 68\% C.L.
\begin{table*}[t]
\centering
\centering
\begingroup
\setlength{\tabcolsep}{5pt}
\renewcommand{\arraystretch}{1.18}
\fontsize{8}{10}\selectfont
\begin{tabular}{@{}lccccc@{}}
\toprule
\addlinespace[2pt]
Parameter & $\Lambda$CDM & $\Lambda$CDM+$\xi$ & $N_{\rm eff}$ & $N_{\rm eff}+\xi$ & $N_{\rm eff}+\delta_q$ \\
\midrule
$H_0$ & $68.01\pm0.26$ & $68.00\pm0.27$ & $68.10\pm0.51$ & $69.24\pm0.92$ & $68.18\pm0.52$ \\
$100\Omega_bh^2$ & $2.225\pm0.010$ & $2.225\pm0.011$ & $2.226\pm0.012$ & $2.238\pm0.014$ & $2.233\pm0.013$ \\
$\Omega_ch^2$ & $0.11785\pm0.00060$ & $0.11785\pm0.00060$ & $0.1181\pm0.0014$ & $0.1215\pm0.0027$ & $0.1180\pm0.0014$ \\
$n_s$ & $0.9678\pm0.0034$ & $0.9677\pm0.0033$ & $0.9682\pm0.0041$ & $0.9726\pm0.0050$ & $0.9684\pm0.0040$ \\
$N_{\rm eff}$ & $3.044$ & $3.044+\Delta N_{\rm eff}^{\xi}$ & $3.058\pm0.075$ & $3.27\pm0.16$ & $3.057\pm0.075$ \\
$\xi$ & $0$ & $0.0033\pm0.0054$ & $0$ & $0.017\pm0.011$ & $0$ \\
$\delta_q$ & $0$ & $0$ & $0$ & $0$ & $0.0056\pm0.0041$ \\
$\Omega_m$ & $0.3043\pm0.0034$ & $0.3044\pm0.0035$ & $0.3040\pm0.0037$ & $0.3015\pm0.0040$ & $0.3032\pm0.0037$ \\
$Y_P^{\rm BBN}$ & $0.246935\pm0.000045$ & $0.2462\pm0.0013$ & $0.2471\pm0.0010$ & $0.2459\pm0.0013$ & $0.2464\pm0.0012$ \\
$10^5\mathrm{D/H}$ & $2.468\pm0.019$ & $2.464\pm0.020$ & $2.470\pm0.024$ & $2.495\pm0.030$ & $2.512\pm0.039$ \\
$\sigma_8$ & $0.8056\pm0.0057$ & $0.8055^{+0.0052}_{-0.0058}$ & $0.8064\pm0.0065$ & $0.8143\pm0.0086$ & $0.8062\pm0.0066$ \\
$S_8$ & $0.8114\pm0.0077$ & $0.8114\pm0.0077$ & $0.8117\pm0.0077$ & $0.8163\pm0.0086$ & $0.8104\pm0.0079$ \\
\midrule[0.9pt]
\addlinespace[2pt]
Parameter & $N_{\rm eff}+\sum m_\nu$ & $N_{\rm eff}+\xi+\sum m_\nu$ & EDE & EDE+$\xi$  & EDE+$\delta_q$ \\
\midrule
$H_0$ & $68.19\pm0.52$ & $69.14\pm0.92$ & $68.50^{+0.25}_{-0.70}$ & $68.45^{+0.24}_{-0.64}$ & $68.85^{+0.33}_{-0.96}$ \\
$100\Omega_bh^2$ & $2.224\pm0.012$ & $2.234\pm0.015$ & $2.229\pm0.011$ & $2.229\pm0.012$ & $2.242^{+0.013}_{-0.016}$ \\
$\Omega_ch^2$ & $0.1182\pm0.0014$ & $0.1210\pm0.0026$ & $0.11961^{+0.00042}_{-0.0024}$ & $0.11950^{+0.00045}_{-0.0022}$ & $0.12035^{+0.00082}_{-0.0032}$ \\
$n_s$ & $0.9670\pm0.0040$ & $0.9709\pm0.0052$ & $0.9700^{+0.0037}_{-0.0053}$ & $0.9694^{+0.0036}_{-0.0051}$ & $0.9717^{+0.0040}_{-0.0070}$ \\
$N_{\rm eff}$ & $3.044\pm0.076$ & $3.22\pm0.16$ & $3.044$ & $3.044+\Delta N_{\rm eff}^{\xi}$ & $3.044$ \\
$\xi$ & $0$ & $0.014\pm0.011$ & $0$ & $0.0032\pm0.0054$ & $0$  \\
$\delta_q$ & $0$ & $0$  & $0$ & $0$ & $0.0072\pm0.0044$ \\
$\sum m_\nu$ & $<0.0640$ & $<0.0777$ & $0.06$ & $0.06$ & $0.06$  \\
$f_{\rm EDE}$ & --- & --- & $<0.0120$ & $<0.0113$ & $<0.0187$ \\
$\Omega_m$ & $0.3026\pm0.0037$ & $0.3006\pm0.0040$ & $0.3038\pm0.0035$ & $0.3040\pm0.0035$ & $0.3026\pm0.0036$ \\
$Y_P^{\rm BBN}$ & $0.2469\pm0.0011$ & $0.2460\pm0.0013$ & $0.246954\pm0.000049$ & $0.2462\pm0.0013$ & $0.24601\pm0.00058$ \\
$10^5\mathrm{D/H}$ & $2.470\pm0.024$ & $2.490\pm0.029$ & $2.460\pm0.021$ & $2.457\pm0.022$ & $2.508\pm0.037$ \\
$\sigma_8$ & $0.8130^{+0.0079}_{-0.0069}$ & $0.8189\pm0.0089$ & $0.8089^{+0.0059}_{-0.0081}$ & $0.8083^{+0.0058}_{-0.0083}$ & $0.8107^{+0.0064}_{-0.010}$ \\
$S_8$ & $0.8165\pm0.0084$ & $0.8196\pm0.0086$ & $0.8141^{+0.0080}_{-0.0095}$ & $0.8136^{+0.0080}_{-0.0095}$ & $0.8141^{+0.0086}_{-0.0106}$ \\
\bottomrule
\end{tabular}
\endgroup

\caption{Marginalized parameter constraints from Planck, DESI DR2 BAO, Pantheon+, and BBN for all ten models. Entries are 68\% credible intervals or upper limits, except for the upper limits on $\sum m_\nu$, which are quoted at 95\% C.L. We define $S_8\equiv\sigma_8(\Omega_m/0.3)^{1/2}$. $H_0$ is in ${\rm km\,s^{-1}\,Mpc^{-1}}$ and $\sum m_\nu$ in eV.}
\label{tab:constraints}
\end{table*}

\subsection{$\Lambda$CDM+$\xi$}
\label{sec:results_LCDMxi}
We first consider lepton asymmetry in the absence of an
independent extra radiation component, setting
$N_{\rm eff}=3.044+\Delta N_{\rm eff}^{\xi}$.
For $\Lambda$CDM+$\xi$, we obtain
\begin{align}
 \xi&=0.0033\pm0.0054,\nonumber\\
 H_0&=68.00\pm0.27\,{\rm km\,s^{-1}\,Mpc^{-1}}.
\end{align}
The degeneracy parameter is consistent with zero, and the
inferred $H_0$ is almost unchanged from
$68.01\pm0.26\,{\rm km\,s^{-1}\,Mpc^{-1}}$ in $\Lambda$CDM.
The baryon and cold dark matter densities, $n_s$, and $\sigma_8$
also remain close to their $\Lambda$CDM values.
Our 95\% constraint,
$\xi=0.003^{+0.011}_{-0.010}$, is consistent with and
of similar precision to $\xi=0.006\pm0.011$ (95\% C.L.)
reported in Ref.~\cite{Ghedini:2026mvj}, based on Planck
data with high-$\ell$ cuts for combination with ACT DR6,
together with SPT-3G D1, DESI DR2, the LBT helium abundance,
and the PDG deuterium constraint.

On the other hand, the derived posterior for the helium
abundance broadens appreciably, from
$Y_P^{\rm BBN}=0.246935\pm0.000045$ in $\Lambda$CDM to $0.2462\pm0.0013$ in $\Lambda$CDM+$\xi$.
In $\Lambda$CDM, the predicted helium abundance depends mainly
on the tightly constrained baryon density.
Allowing $\xi$ to vary introduces an additional dependence
through the neutron--proton conversion rates.
A positive $\xi$ lowers the neutron-to-proton ratio and hence
reduces $Y_P^{\rm BBN}$, producing the strong anticorrelation
shown in Fig.~\ref{fig:lcdmxi}.
The narrow helium interval in $\Lambda$CDM therefore reflects
the assumed BBN model and the baryon-density constraint,
rather than an independent measurement of the helium abundance
at this precision. The quoted intervals reflect the variation of the BBN predictions
with the sampled cosmological parameters; the BBN theoretical
uncertainties are included in the likelihood but are not added
to the reported abundance intervals.
Deuterium remains nearly unchanged,
$10^5\mathrm{D/H}=2.464\pm0.020$, and retains its strong
anticorrelation with the baryon density.

\subsection{$\Neff+\xi$}
\label{sec:results_neffxi}
When $N_{\rm eff}$ is varied with $\xi=0$, we obtain
\begin{align}
 N_{\rm eff}&=3.058\pm0.075,\nonumber\\
 H_0&=68.10\pm0.51\,{\rm km\,s^{-1}\,Mpc^{-1}}.
\end{align}
The effective number of relativistic species is consistent
with the standard value, with
$N_{\rm eff}-3.044=0.014\pm0.075$.
For comparison, using the same baseline CMB, BAO, and SNe data
but removing the BBN likelihoods, we obtain
$N_{\rm eff}=3.21\pm0.18$.\footnote{This result is consistent with the DESI DR2 BAO+CMB
analysis of Ref.~\cite{Elbers:2025vlz}, which found
$N_{\rm eff}=3.23\pm0.18$ at 68\% C.L. using
Planck PR4, Planck/ACT lensing, and DESI DR2 BAO without
primordial-abundance measurements.}
Including BBN therefore shifts the posterior toward the standard
value and reduces its uncertainty by more than a factor of two.
The primordial-helium measurement contributes to this tightening:
a larger radiation density increases the expansion rate during BBN,
leading to earlier weak freeze-out and a larger neutron abundance,
and hence to a larger predicted $Y_P$.
The LBT helium measurement therefore disfavors the larger-$N_{\rm eff}$
region allowed by the CMB, BAO, and SNe data alone.
The central $H_0$ correspondingly decreases from $69.0\pm1.0$
to $68.10\pm0.51\,{\rm km\,s^{-1}\,Mpc^{-1}}$.

The situation changes when both $N_{\rm eff}$ and $\xi$ are varied. We find
\begin{equation}
\begin{aligned}
 N_{\rm eff} &= 3.27\pm0.16,\\
 \xi         &= 0.017\pm0.011,\\
 H_0         &= 69.24\pm0.92\,{\rm km\,s^{-1}\,Mpc^{-1}}.
\end{aligned}
\label{eq:neffxi_constraints}
\end{equation}
Compared with the model with only $N_{\rm eff}$ varied, the
central $H_0$ increases by $1.14\,{\rm km\,s^{-1}\,Mpc^{-1}}$,
while its allowed range also broadens.
A positive $\xi$ reduces the neutron-to-proton ratio and
compensates the helium increase induced by extra radiation.
The positive correlations in Fig.~\ref{fig:neffxi} illustrate
this compensation, which permits larger $N_{\rm eff}$ and $H_0$
while maintaining consistency with the LBT helium measurement.
The resulting helium posterior is $Y_P^{\rm BBN}=0.2459\pm0.0013$.
Meanwhile, the deuterium abundance increases from
$10^5\mathrm{D/H}=2.470\pm0.024$ to $2.495\pm0.030$,
closer to the adopted measurement, $2.508\pm0.029$.
The baryon density increases from $\Omega_bh^2=0.02226\pm0.00012$
to $0.02238\pm0.00014$, partly counteracting the deuterium
increase induced by the faster expansion during BBN
at larger $N_{\rm eff}$.

The radiation density generated directly by $\xi$ remains much
smaller than the inferred excess in $N_{\rm eff}$.
At $\xi=0.017$, the radiation-density correction defined in
Sec.~\ref{sec:models} gives
$\Delta N_{\rm eff}^{\xi}\simeq3.8\times10^{-4}$,
whereas the central excess in $N_{\rm eff}$ is approximately $0.23$.
The main role of the asymmetry is therefore to modify the weak
rates during BBN, allowing additional radiation.
For comparison, Ref.~\cite{Li:2025rjr} obtained
$\xi=0.056\pm0.017$ and $\delta N_{\rm eff}=0.41\pm0.16$
using Planck, DESI DR1, Pantheon+, EMPRESS helium,
and PDG deuterium data.
Although our central values are smaller, both $\xi$ and
$N_{\rm eff}-3.044$ remain positive with the updated datasets
and the LBT helium abundance, which is consistent with
standard BBN.
Thus, a helium measurement below the standard BBN prediction
is not necessary for this compensation mechanism to permit
additional radiation and a higher inferred $H_0$.

As a comparison, Table~\ref{tab:constraints} also includes
$N_{\rm eff}+\delta_q$, for which $N_{\rm eff}=3.057\pm0.075$,
$\delta_q=0.0056\pm0.0041$, and
$H_0=68.18\pm0.52\,{\rm km\,s^{-1}\,Mpc^{-1}}$.
Unlike lepton asymmetry, the allowed quark-mass variation leaves
the constraints on $N_{\rm eff}$ and $H_0$ nearly unchanged.

\subsection{Constraints with sampled neutrino mass}
\label{sec:results_massive}
We now sample $\sum m_\nu$ in $N_{\rm eff}+\sum m_\nu$ and
$N_{\rm eff}+\xi+\sum m_\nu$. The results are collected in
Table~\ref{tab:constraints} and Fig.~\ref{fig:neffmnu}.
Sampling the mass leaves the radiation and asymmetry constraints
largely unchanged. We find $N_{\rm eff}=3.044\pm0.076$ for
$N_{\rm eff}+\sum m_\nu$, and
\begin{equation}
 N_{\rm eff}=3.22\pm0.16,\qquad \xi=0.014\pm0.011
\end{equation}
when both $\xi$ and the mass are sampled. Similarly, the inferred
Hubble constants, $H_0=68.19\pm0.52$ and $69.14\pm0.92$\,
${\rm km\,s^{-1}\,Mpc^{-1}}$, are close to their fixed-mass
counterparts.

For the summed neutrino mass, we obtain
\begin{equation}
 \sum m_\nu<\begin{cases}
   0.0640\,{\rm eV} & (N_{\rm eff}+\sum m_\nu),\\
   0.0777\,{\rm eV} & (N_{\rm eff}+\xi+\sum m_\nu),
 \end{cases}
 \quad(95\%\ \mathrm{C.L.}).
 \label{eq:mnu_limits}
\end{equation}
Allowing a common lepton asymmetry therefore relaxes the upper
limit by $0.0137\,{\rm eV}$, or about $21\%$.
Both models favor small neutrino masses, yielding only upper limits.
Sampling the mass modestly raises $\sigma_8$ and $S_8$
(Table~\ref{tab:constraints}), consistent with weaker neutrino
free-streaming suppression when masses below the fiducial
$0.06\,{\rm eV}$ are allowed. 

\subsection{Axionlike EDE}
\label{sec:results_ede}

We first examine how primordial-abundance measurements constrain
the EDE model. Without the BBN likelihoods, we obtain
\begin{align}
 f_{\rm EDE}&<0.0171, &
 H_0&=68.80^{+0.30}_{-0.92}\;
 {\rm km\,s^{-1}\,Mpc^{-1}},
\end{align}
where the upper limit and uncertainties are at 68\% C.L.
The corresponding baryon density and predicted deuterium abundance
are $\Omega_bh^2=0.02241^{+0.00014}_{-0.00016}$ and
$10^5\mathrm{D/H}=2.440^{+0.028}_{-0.025}$.
A larger EDE contribution can be accompanied by a higher baryon
density, which lowers the predicted primordial deuterium abundance.
The deuterium measurement therefore constrains the parameter
region that permits larger EDE fractions and higher
$H_0$~\cite{Seto:2021xua,Takahashi:2023twt, Bella:2026zuk}.
Including the primordial-abundance likelihoods gives
\begin{align}
 f_{\rm EDE}&<0.0120, &
 H_0&=68.50^{+0.25}_{-0.70}\;
 {\rm km\,s^{-1}\,Mpc^{-1}},
\end{align}
as summarized in Table~\ref{tab:constraints}.
Thus, BBN tightens the upper limit on $f_{\rm EDE}$ from
$0.0171$ to $0.0120$ and lowers the central $H_0$ by
$0.30\,{\rm km\,s^{-1}\,Mpc^{-1}}$.
The baryon density decreases to
$\Omega_bh^2=0.02229\pm0.00011$, raising the predicted
deuterium abundance to $10^5\mathrm{D/H}=2.460\pm0.021$,
closer to the observed value.

Allowing lepton asymmetry does not relax this BBN restriction.
For EDE+$\xi$, we find
\begin{align}
 f_{\rm EDE}&<0.0113, &
 H_0&=68.45^{+0.24}_{-0.64}\;
 {\rm km\,s^{-1}\,Mpc^{-1}},
\end{align}
with $\xi=0.0032\pm0.0054$, consistent with zero.
The baryon density, $\Omega_bh^2=0.02229\pm0.00012$,
and deuterium abundance, $10^5\mathrm{D/H}=2.457\pm0.022$,
also remain nearly unchanged.
The helium measurement restricts $\xi$ to values whose effect
on deuterium is too small to compensate for the decrease
associated with a higher baryon density.
Moreover, a small positive $\xi$ reduces D/H in the adopted prescription,
rather than compensating this decrease.

We next allow the light-quark mass during BBN to vary according to
Sec.~\ref{sec:mq}. For EDE+$\delta_q$, we obtain
\begin{equation}
\begin{aligned}
\delta_q &= 0.0072\pm0.0044,\\
f_{\rm EDE} &< 0.0187,\\
H_0 &= 68.85^{+0.33}_{-0.96}\,{\rm km\,s^{-1}\,Mpc^{-1}},
\end{aligned}
\label{eq:edemq_constraints}
\end{equation}
where the intervals and the upper limit are at 68\% C.L.
The positive mean corresponds to a $0.72\%$ light-quark-mass
increase during BBN, with a displacement of approximately
$1.6\sigma$ from zero in the Gaussian approximation.
Compared with EDE at fixed quark mass, the upper limit
on $f_{\rm EDE}$ increases from $0.0120$ to $0.0187$,
and the central value of $H_0$ increases by
$0.35\,{\rm km\,s^{-1}\,Mpc^{-1}}$.
The resulting $f_{\rm EDE}$ upper limit and $H_0$ constraint
are close to those obtained for EDE without the BBN likelihoods,
although the posterior still yields only an upper limit on the EDE fraction.
Figure~\ref{fig:edemq} illustrates the broader allowed ranges
of $f_{\rm EDE}$ and $H_0$.

The inferred baryon density increases to
$\Omega_bh^2=0.02242^{+0.00013}_{-0.00016}$, compared with
$0.02229\pm0.00011$ in EDE. We find
\begin{equation}
\begin{aligned}
Y_P^{\rm BBN} &= 0.24601\pm0.00058,\\
10^5\mathrm{D/H} &= 2.508\pm0.037,
\end{aligned}
\end{equation}
consistent with the adopted measurements.
The central deuterium abundance increases from $2.460$ in EDE
to $2.508$ in EDE+$\delta_q$ despite the larger baryon density.
This reflects the positive response of deuterium to
$\delta_q$ in Eq.~\ref{eq:mqDH}.
At $\delta_q=0.0072$, the adopted response factors increase
D/H by approximately $2.8\%$ and decrease $Y_P$ by
approximately $0.40\%$ at fixed $(\omega_b,N_{\rm eff})$.
The matter fluctuation amplitude is
$\sigma_8=0.8107^{+0.0064}_{-0.010}$, while
$S_8=0.8141^{+0.0086}_{-0.0106}$ is close to that of EDE.

\section{Discussion}
\label{sec:discussion}
To quantify the remaining Hubble tension in each model,
we compare the inferred Hubble constants with the
updated SH0ES measurement,
$H_0^{\rm SH0ES}=73.17\pm0.86\,{\rm km\,s^{-1}\,Mpc^{-1}}$
\cite{Breuval:2024SMC}, using the Gaussian tension
\begin{equation}
 T_{H_0}=\frac{|H_0^{\rm model}-H_0^{\rm SH0ES}|}
 {\sqrt{\sigma_{H_0,\rm model}^2+\sigma_{H_0,\rm SH0ES}^2}}.
 \label{eq:gaussian_tension}
\end{equation}

\begin{table}[t]
\centering
\begingroup
\small
\setlength{\tabcolsep}{4pt}
\renewcommand{\arraystretch}{1.09}
\begin{tabular}{lcc}
\toprule
Model & $H_0$ & $T_{H_0}$ \\
\midrule
$\Lambda$CDM                  & $68.01\pm0.26$ & $5.74\sigma$ \\
$\Lambda$CDM+$\xi$           & $68.00\pm0.27$ & $5.74\sigma$ \\
$N_{\rm eff}$                  & $68.10\pm0.51$ & $5.07\sigma$ \\
$N_{\rm eff}+\xi$              & $69.24\pm0.92$ & $3.12\sigma$ \\
$N_{\rm eff}+\delta_q$         & $68.18\pm0.52$ & $4.97\sigma$ \\
$N_{\rm eff}+\sum m_\nu$       & $68.19\pm0.52$ & $4.96\sigma$ \\
$N_{\rm eff}+\xi+\sum m_\nu$  & $69.14\pm0.92$ & $3.20\sigma$ \\
EDE                              & $68.50^{+0.25}_{-0.70}$ & $4.75\sigma$ \\
EDE+$\xi$                     & $68.45^{+0.24}_{-0.64}$ & $4.89\sigma$ \\
EDE+$\delta_q$                & $68.85^{+0.33}_{-0.96}$ & $4.02\sigma$ \\
\bottomrule
\end{tabular}
\endgroup
\caption{Gaussian $H_0$ discrepancies with SH0ES
($73.17\pm0.86\,{\rm km\,s^{-1}\,Mpc^{-1}}$).
All cosmological fits include BBN. For the EDE models, asymmetric
68\% uncertainties are symmetrized by averaging the upper and lower errors.}
\label{tab:h0_tension}
\end{table}

Table~\ref{tab:h0_tension} shows that the smallest Gaussian
$H_0$ discrepancies are obtained with $N_{\rm eff}+\xi$: $3.12\sigma$ at fixed neutrino mass and $3.20\sigma$
when the mass is sampled, compared with $5.74\sigma$ in $\Lambda$CDM.
This reduction reflects both the upward shift in the central $H_0$
and the broadening of its posterior uncertainty.
EDE+$\delta_q$ gives $4.02\sigma$, improving on EDE and
EDE+$\xi$, while $N_{\rm eff}+\delta_q$ remains close to
$N_{\rm eff}$ alone. 

Our results for extra radiation and EDE show that the effectiveness of lepton asymmetry and quark-mass variation depends on how BBN constrains each model. Extra radiation increases the expansion rate during BBN, enhancing helium production; a positive $\xi$ compensates this effect and thereby allows larger $N_{\rm eff}$. EDE instead shifts the baryon density preferred by cosmological data toward higher values, reducing the predicted deuterium abundance; a positive $\delta_q$ compensates this effect and thereby allows larger EDE fractions. Adding $\delta_q$, however, leaves the $N_{\rm eff}$ constraint nearly unchanged.

Our results may also have implications for cosmic birefringence,
a rotation of the CMB polarization plane for which evidence
has been reported at $2.4\sigma$~\cite{Minami:2020odp}
and subsequently at $3.6\sigma$~\cite{Eskilt:2022cff}.
Such a rotation can be sourced by a neutrino-asymmetry current
through an effective Chern--Simons interaction~\cite{Yin:2026qlh},
with its amplitude depending on $b_J\xi_{\nu_e}/M^2$,
where $b_J$ is a dimensionless coupling and $M$ is the
interaction scale.
The positive lepton asymmetry favored by our
$N_{\rm eff}+\xi$ analysis could therefore also account for
the reported birefringence signal for suitable values of
$b_J$ and $M$.

The interpretation of the quark-mass results remains conditional
on the nuclear response in Eqs.~\ref{eq:mqYp} and \ref{eq:mqDH}.
The modest displacement of $\delta_q$ from zero suggests that
efforts to alleviate the Hubble tension within EDE models may
also benefit from small modifications to BBN nuclear physics.
The response coefficients are fixed to their central estimates
in our analysis; propagating their uncertainties into the
cosmological constraints remains a task for future work.
Quark-mass variation may also have implications for the primordial
lithium problem, although its effect depends on the nuclear
response, particularly the shifts of resonance
energies~\cite{Berengut:2009js,Cheoun:2011yn}.
Under suitable assumptions about these shifts, quark-mass variation
has been reported to bring the predicted lithium abundance into
agreement with observations~\cite{Mori:2019cfo}.

\section{Conclusions}
\label{sec:conclusions}
Extra radiation and early dark energy can reduce the sound
horizon before recombination and increase the inferred
Hubble constant.
However, their compatibility with primordial light-element
abundances must also be examined.
We have compared these scenarios with and without primordial
lepton asymmetry, using Planck CMB, DESI DR2 BAO, Pantheon+,
and updated helium and deuterium measurements.
In particular, we have examined whether lepton asymmetry
can relax the BBN constraints when the precise LBT helium
measurement, which agrees with standard BBN, is adopted.
For EDE, we have also examined light-quark-mass variation
during BBN as a separate mechanism for modifying
the primordial abundances.

In the extra radiation model without lepton asymmetry,
including the BBN likelihoods changes the constraint from
$N_{\rm eff}=3.21\pm0.18$ to $3.058\pm0.075$.
The latter is consistent with the standard value and
strongly restricts the increase in the inferred Hubble
constant.
When both $N_{\rm eff}$ and $\xi$ are varied, we obtain
\begin{equation}
\begin{aligned}
N_{\rm eff} &= 3.27\pm0.16,\\
\xi &= 0.017\pm0.011,\\
H_0 &= 69.24\pm0.92\,{\rm km\,s^{-1}\,Mpc^{-1}}.
\end{aligned}
\end{equation}
A positive $\xi$ reduces the neutron-to-proton ratio and
compensates the increase in primordial helium caused by
extra radiation.
The Gaussian discrepancy with SH0ES is reduced from
$5.74\sigma$ in $\Lambda$CDM to $3.12\sigma$.
We therefore find that the extra radiation model with
lepton asymmetry can still alleviate the Hubble tension
without adopting a helium abundance below the standard
BBN prediction.
When the summed active-neutrino mass is also varied, allowing
lepton asymmetry relaxes its 95\% upper limit from 0.0640 to
0.0777 eV, while the constraints on $N_{\rm eff}$, $\xi$, and
$H_0$ remain close to those obtained at fixed mass.

For the axionlike EDE model considered here, we find a
different result.
Including the BBN likelihoods tightens the 68\% upper limit
on $f_{\rm EDE}$ from $0.0171$ to $0.0120$ and changes
$H_0$ from $68.80^{+0.30}_{-0.92}$ to
$68.50^{+0.25}_{-0.70}\,{\rm km\,s^{-1}\,Mpc^{-1}}$.
Allowing lepton asymmetry gives $f_{\rm EDE}<0.0113$
at 68\% C.L. and
$H_0=68.45^{+0.24}_{-0.64}\,{\rm km\,s^{-1}\,Mpc^{-1}}$,
leaving the inferred Hubble constant nearly unchanged.
Although the EDE contribution is negligible during BBN,
the larger baryon density associated with a larger EDE
contribution reduces the predicted deuterium abundance.
Our results show that allowing lepton asymmetry does not
effectively relax this restriction: the values of $\xi$ allowed
by the helium measurement have little effect on deuterium,
and a small positive $\xi$ slightly lowers its abundance
in the adopted prescription.

Allowing light-quark-mass variation during BBN gives a
different result from allowing lepton asymmetry in EDE.
For EDE+$\delta_q$, we obtain
\begin{equation}
\begin{aligned}
\delta_q &= 0.0072\pm0.0044,\\
f_{\rm EDE} &< 0.0187,\\
H_0 &= 68.85^{+0.33}_{-0.96}\,{\rm km\,s^{-1}\,Mpc^{-1}},
\end{aligned}
\end{equation}
where the intervals and the upper limit are at 68\% C.L.
The baryon density increases to
$\Omega_bh^2=0.02242^{+0.00013}_{-0.00016}$, while the predicted
abundances, $Y_P^{\rm BBN}=0.24601\pm0.00058$ and
$10^5\mathrm{D/H}=2.508\pm0.037$, remain consistent
with the adopted measurements.
In the nuclear-response prescription used here, a positive
$\delta_q$ increases deuterium and slightly reduces helium.
It can therefore compensate the deuterium decrease caused
by a larger baryon density and relax the BBN constraint
on the EDE fraction.
The Gaussian discrepancy with SH0ES decreases to
approximately $4.02\sigma$, compared with $5.74\sigma$
in $\Lambda$CDM.
These constraints use the central nuclear-response coefficients
without marginalizing over their theoretical uncertainties.
These results suggest that efforts to alleviate the Hubble
tension within EDE models may also benefit from small
modifications to BBN nuclear physics.

Among the models studied, extra radiation with lepton
asymmetry gives the highest central value of the Hubble constant
and the smallest Gaussian discrepancy with SH0ES.
The comparison shows that the ability to relax BBN
constraints depends on how each extension changes the
primordial abundances.
Lepton asymmetry permits additional radiation by
compensating its effect on helium, but does not relax
the deuterium constraint on EDE.
Light-quark-mass variation can instead compensate the
deuterium decrease associated with a larger baryon density,
allowing a larger EDE contribution.
Precision measurements of both abundances, together with
improved calculations of their nuclear response, are therefore
important for assessing these approaches to alleviating
the Hubble tension.

\section*{Acknowledgments}
 YT is supported by JSPS KAKENHI Grant No. 26K17154.

\bibliographystyle{elsarticle-num}
\bibliography{references}

\clearpage

\end{document}